\documentstyle[12pt]{article}

\begin{document}
\addtolength{\baselineskip}{5pt}

\begin{flushright}
hep-th/9509162\\
KUCP-0083\\
September 1995
\end{flushright}

\begin{center}
\begin{Large}
     {\bf Whitham-Toda Hierarchy }\\[3mm]
        {\bf  And } \\[3mm]
      {\bf N = 2 Supersymmetric Yang-Mills Theory}
\end{Large}

\vspace{25pt}

\noindent
Toshio NAKATSU \raisebox{2mm}{{\footnotesize 1}}
and
Kanehisa TAKASAKI \raisebox{2mm}{{\footnotesize 2}{$\dagger$}}
\vspace{18pt}

\begin{small}
$~^{1}${\it Department of Mathematics and Physics, Ritsumeikan University }\\
       {\it Kusatsu,Shiga 525-77,Japan}\\
$~^{2}${\it Department of Fundamental Sciences,
            Faculty of Integrated Human Studies, Kyoto University }\\
       {\it Yoshida-Nihonmatsu-cho,Sakyo-ku,Kyoto 606,Japan}\\
\end{small}

\vspace{30pt}

\underline{ABSTRACT}
\end{center}
\vspace{10pt}
\begin{small}

The exact solution of $N=2$ supersymmetric $SU(N)$ Yang-Mills
theory is studied in the framework of the Whitham hierarchies.
The solution is identified with a homogeneous solution of a
Whitham hierarchy. This integrable hierarchy (Whitham-Toda
hierarchy) describes modulation of a quasi-periodic solution
of the (generalized) Toda lattice hierarchy associated with the
hyperelliptic curves over the quantum moduli space. The relation
between the holomorphic pre-potential of the low energy
effective action  and the $\tau$ function of the (generalized)
Toda lattice hierarchy is also clarified.

\end{small}

\vfill
\hrule
\vskip 3mm
\begin{small}
\noindent{$\dagger$}
E-mail: takasaki@yukawa.kyoto-u.ac.jp, nakatsu@tkyvax.phys.s.u-tokyo.ac.jp
\end{small}

\newpage

    Recently Seiberg and Witten \cite{SW} obtained exact expressions for
the metric on the moduli space and the dyon spectrum of $N=2$ supersymmetric
$SU(2)$ Yang-Mills theory by using a version of the Olive-Montonen duality
\cite{duality} and  holomorphy \cite{seiberg}
of 4$d$ supersymmetric theories.
Their approach has been generalized to the case of other Lie group
\cite{N=2SUSY},\cite{N=2SUSY2}.
Especially surprising in these results is
unexpected emergence of elliptic (or hyperelliptic) curves and their periods.
Although these objects appear
in  the course of determining the holomorphic pre-potential $\cal F$
of the exact low energy effective actions,
physical significance of the curves themselves is unclear yet.
It will be important to clarify their physical role.
An interesting step in this direction has been taken
\cite{integrable} from the view of integrable systems,
in which the correspondence between the Seiberg-Witten solution \cite{SW}
and the Gurevich-Pitaevsky solution \cite{GP}
to the elliptic Whitham-KdV equation \cite{Whitham}
is pointed out.

            In this article we shall consider
the exact solution of $N=2$ supersymmetric $SU(N)$
Yang-Mills theory in the framework of the Whitham hierarchies
\cite{Whitham hierarchy2} and show that it can be identified
with a homogeneous solution of a Whitham hierarchy. This
Whitham hierarchy (Whitham-Toda hierarchy) turns out to
be modulation equations \cite{D-L} of a quasi-periodic
solution of the (generalized) Toda lattice hierarchy
associated with the hyperelliptic curves over the moduli space
of Yang-Mills theory. In particular  the relation between
the pre-potential of this supersymmetric theory and the
$\tau$ function of the (generalized) Toda lattice hierarchy
will be clarified.

\section{}

      One of the main ingredients in the analysis \cite{SW},
\cite{N=2SUSY} of the exact solution of
$N=2$ supersymmetric $SU(N)$ Yang-Mills theory is
the meromorphic differential $dS$
\begin{eqnarray}
dS = \frac{x \frac{dP(x)}{dx}}{y}dx~~~~
\label{dS-1}
\end{eqnarray}
on the family of the hyperelliptic curves
\begin{eqnarray}
C:\ y^2 = P(x)^2 - \Lambda^{2N}~,~~~~~~
 P(x)=x^N+\sum_{k=0}^{N-2}u_{N-k}x^k .
\label{curveC}
\end{eqnarray}
Each $u_k$ is an order parameter of $N=2$ supersymmetric $SU(N)$
Yang-Mills theory.
$u=(u_2, \cdots, u_{N})$ become the parameters of the flat moduli.
$\Lambda $ is the lambda-parameter of this gauge theory and
we henceforth fix its value.
The spectrum of excitations in the theory will be measured by the units
\begin{eqnarray}
a_i=\oint_{\alpha_i}dS~~~,~~~a_{D,i}=\oint_{\beta_i}dS~~~(1 \leq i \leq N-1),
\label{a-aD}
\end{eqnarray}
where $\alpha_i$ and $\beta_i$$(1 \leq i \leq N-1)$ are
the standard symplectic basis of homology cycles of the
hyperelliptic curve $C$. In particular each $\alpha_i$ is
a cycle which encircles counterclockwise the cut between
two neighboring branch points in $x-$plane.
Let $p_{\infty}$ and $\tilde{p}_{\infty}$ be
the two points at infinity of the Riemann sheets of $C$ :
$x(p_{\infty})=x(\tilde{p}_{\infty})=\infty. $
Since the pole divisor of  $dS$ (\ref{dS-1}) is
$2p_{\infty}+2\tilde{p}_{\infty}$
and it has no residues at these points,
one can decompose it into the sum :
\begin{eqnarray}
dS=dX_{\infty,1}
+\sum_{i=1}^{N-1}a_idz_i~~~,
\label{decomposition of dS-1}
\end{eqnarray}
where $dz_i (1 \leq i \leq N-1)$ is the holomorphic differential
normalized by the condition $\oint_{\alpha_i}dz_j=\delta_{i,j}$.
$dX_{\infty,1}$ is the meromorphic differential
of second kind (its second order poles are at
$p_{\infty}$ and $\tilde{p}_{\infty}$)
with vanishing periods along the
$\alpha$-cycles, that is,
$\oint_{\alpha_i}dX_{\infty,1}=0~~$ for $~^{\forall}i$.

    Let us investigate the role of $dS$ from the view of integrable system.
For this purpose we introduce the following meromorphic functions
$h$ and $\tilde{h}$ on the curve $C$,
\begin{eqnarray}
h=y+P(x)~~~,~~~\tilde{h}=-y+P(x)~~~,
\label{def of h}
\end{eqnarray}
and then consider the effect of infinitesimal deformation of the
moduli parameters $u$ with $h$ (or $\bar{h}$) being fixed.
Under this condition one can obtain
\begin{eqnarray}
\left. \frac{\partial}{\partial u_{N-k}}dS~~ \right|_{fix~h}
=-\frac{x^k}{y}dx~~.
\end{eqnarray}
After changing the moduli parameters from $u$ to $a=(a_1, \cdots, a_{N-1})$
the above equation reads
\begin{eqnarray}
\left. \frac{\partial}{\partial a_i}dS ~~\right|_{fix~h}
= dz_i~~~,
\label{del-a of dS-1}
\end{eqnarray}
which implis that $dS$ satisfies the following system of
differential equations:
\begin{eqnarray}
\frac{\partial}{\partial a_i} dz_j =
\frac{\partial}{\partial a_j} dz_i~~~~~(1 \leq i,j \leq N-1),
\label{FFM-1}
\end{eqnarray}
where the derivation by the moduli parameters $a$ is understood
to be partial derivation fixing $h$ and the other $a$'s.
      Differential equations of this form (\ref{FFM-1})
were first derived by Flashka, Forest and MacLaughlin
\cite{Whitham hierarchy} as modulation equations
of quasi-periodic solutions in soliton theory.
The concept of modulated quasi-periodic solutions
originates in Whitham's work on the KdV equation \cite{Whitham},
and because of this, this type of systems are called
Whitham equations.  Remarkably, Whitham equations themselves
are integrable systems. In particular, as soliton equations
admit infinitely many commuting flows ( which constitute
an integrable hierarchy ), the associated Whitham equations
also have infinitely many extra commuting flows
\cite{Whitham hierarchy2}.  These commuting flows are
generated by a set of meromorphic differentials
$ \{ d\Omega_{A} \}_{A \in I}$,  where flows are
labeled by indices $A,B,\cdots$, and the hierarchy
can be written as integrability conditions of those
differentials. In the above case, these equations are
given by
\begin{eqnarray}
\frac{\partial}{\partial a_i} dz_j  =
\frac{\partial}{\partial a_j} dz_i~~,~~
\frac{\partial}{\partial T_A} dz_i  =
\frac{\partial}{\partial a_i} d\Omega_A~~,~~
\frac{\partial}{\partial T_A} d\Omega_B  =
\frac{\partial}{\partial T_B} d\Omega_A~~,
\label{FFM-2}
\end{eqnarray}
where, as in (\ref{FFM-1}), the derivation by the time
variables mean derivation fixing $h$.

\section{}

            Now we specify the meromorphic differentials $d\Omega_A$.
They consist of two types of meromorphic differentials
$d\Omega_{\infty,n}$ and $d\tilde{\Omega}_{\infty,n}$
of second-kind $(n \geq 1)$:
$d\Omega_{\infty,n}$ ($d\tilde{\Omega}_{\infty,n}$)
has a pole of order $n+1$ at $p_{\infty}$ ($\tilde{p}_{\infty}$) and is
holomorphic elsewhere. We also introduce a meromorphic differential
$d\Omega_{\infty,0} (\equiv d\tilde{\Omega}_{\infty,0})$ of third-kind :
$d\Omega_{\infty,0}$ has simple poles at
$p_{\infty}$ and $\tilde{p}_{\infty}$ with
$res_{p_{\infty}}d\Omega_{\infty,0}
=-res_{\tilde{p}_{\infty}}d\Omega_{\infty,0}=1$
and is holomorphic elsewhere.
All these differentials are normalized such that
they have no periods along any $\alpha_i$-cycle ;
$\oint_{\alpha_i} d\Omega_{\infty,n}=
\oint_{\alpha_i} d\tilde{\Omega}_{\infty,n}=0$ for $\forall i$,
and determined by $h$ and $\tilde{h}$
through the following prescription. Let us define local coordinates
$z_{\infty}$ and $\tilde{z}_{\infty}$
in  neighborhoods of
$p_{\infty}$ and $\tilde{p}_{\infty}$ as
\begin{eqnarray}
z_{\infty}^N = h^{-1}~~~,~~~
\tilde{z}_{\infty}^N= \tilde{h}^{-1}~~~.
\label{coordinate at infty}
\end{eqnarray}
Because the divisors of $h$ and $\tilde{h}$ are respectively
$N\tilde{p}_{\infty}-Np_{\infty}$ and $Np_{\infty}-N\tilde{p}_{\infty}$
\footnote{This follows from the relation:
$h \tilde{h}=\Lambda^{2N}$.} it follows that
$z_{\infty}(p_{\infty})=0$ and $\tilde{z}_{\infty}(\tilde{p}_{\infty})=0$.
In a neighborhood of $p_{\infty}$,
they can be written
\begin{eqnarray}
d\Omega_{\infty,n} &=&
\{-nz_{\infty}^{-n-1}-\sum_{m \geq 1}q_{m,n}z_{\infty}^{m-1}\}
dz_{\infty}~~~~~(~n\geq 1~),
\nonumber \\
d\tilde{\Omega}_{\infty,n} &=&
\{\delta_{n,0}z_{\infty}^{-1}-\sum_{m \geq 1}r_{m,n}z_{\infty}^{m-1}\}
dz_{\infty}~~~~~(~n \geq 0~) ,
\label{expansion-1}
\end{eqnarray}
while, in a neighborhood of $\tilde{p}_{\infty}$,
\begin{eqnarray}
d\Omega_{\infty,n} &=&
\{-\delta_{n,0}\tilde{z}_{\infty}^{-1}
-\sum_{m \geq 1}\bar{r}_{n,m}\tilde{z}_{\infty}^{m-1}\}
d\tilde{z}_{\infty}~~~~~(~n \geq 0~) ,
\nonumber \\
d\tilde{\Omega}_{\infty,n} &=&
\{-n\tilde{z}_{\infty}^{-n-1}
-\sum_{m \geq 1}\bar{q}_{m,n}\tilde{z}_{\infty}^{m-1}\}
d\tilde{z}_{\infty}~~~~~(~n\geq 1~).
\label{expansion-2}
\end{eqnarray}

           Integrability conditions (\ref{FFM-2}) clearly
ensure the existence of  a differential $dS$ which satisfies
\begin{eqnarray}
\frac{\partial}{\partial a_i}dS = dz_i,~~
\frac{\partial}{\partial T_n}dS =d\Omega_{\infty,n},~~
\frac{\partial}{\partial \bar{T}_n}dS =d\tilde{\Omega}_{\infty,n},~~
\frac{\partial}{\partial T_0}dS =d\Omega_{\infty,0},
\label{dS-2}
\end{eqnarray}
where $1 \leq i \leq N-1$ and $n \geq 1$.
We denote time variables  of the flows
generated by $d\Omega_{\infty,n}$, $d\Omega_{\infty,0}$
and $d\tilde{\Omega}_{\infty,n}$
as $T_n$, $T_0$ and $\bar{T}_n$ respectively.
Now let us construct  a function
$F(a,T,\overline{T})$ from this differential $dS$
by the following equations
\begin{eqnarray}
\frac{\partial F}{\partial a_i}&=&
\frac{1}{2 \pi \sqrt{-1}}\oint_{\beta_i}dS ~~~~
(~\equiv \frac{a_{D,i}}{2 \pi \sqrt{-1}}~),
\nonumber \\
\frac{\partial F}{\partial T_n}&=&
-res_{p_{\infty}}z_{\infty}^{-n}dS~,~~~~~
\frac{\partial F}{\partial \overline{T}_n}=
-res_{\tilde{p}_{\infty}}\tilde{z}_{\infty}^{-n}dS~~,
\nonumber \\
\frac{\partial F}{\partial T_0}&=&
-res_{p_{\infty}}\ln z_{\infty} dS
+res_{\tilde{p}_{\infty}}\ln \tilde{z}_{\infty} dS~~,
\label{F}
\end{eqnarray}
where $1 \leq i \leq N-1$ and $n \geq 1$.
Notice that  consistency of this definition of the $F$-function
can be checked   by using integrability conditions (\ref{FFM-2})
and the Riemann bilinear relation for meromorphic differentials
\cite{F-K}. As an example let us prove the equality:
\begin{eqnarray}
\frac{\partial }{\partial T_n} \left(
\frac{1}{2 \pi \sqrt{-1}} \oint_{\beta_i}dS \right)
=
\frac{\partial }{\partial a_i} \left(
-res_{p_{\infty}}z_{\infty}^{-n}dS \right)~~~~~(n \geq 1).
\label{example}
\end{eqnarray}
In fact, by setting
$\Omega_{\infty,n}(p)=\int^{p}d\Omega_{\infty,n}$,
the L.H.S of eq.(\ref{example}) can be transformed as
\begin{eqnarray*}
\frac{\partial }{\partial T_n} \left(
\frac{1}{2 \pi \sqrt{-1}} \oint_{\beta_i}dS
\right) &=&
\frac{1}{2 \pi \sqrt{-1}}  \oint_{\beta_i}d\Omega_{\infty,n}
\nonumber \\
&=&
\frac{-1}{2 \pi \sqrt{-1}} \sum_{j=1}^{N-1}\left\{
\oint_{\alpha_j}d\Omega_{\infty,n}
\oint_{\beta_j}dz_i
-
\oint_{\alpha_j}dz_i
\oint_{\beta_j}d\Omega_{\infty,n}\right\}
\nonumber \\
&=&
-res_{p_{\infty}}\Omega_{\infty,n}dz_i
\nonumber \\
&=&
-res_{p_{\infty}}z_{\infty}^{-n}dz_i ,
\end{eqnarray*}
which is equal to the R.H.S of eq.(\ref{example}).
We also notice that, owing to definition (\ref{F}),
the local behaviors of $dS$
can be
described by the $F$-function  as
\begin{eqnarray}
dS &=& \left\{
-\sum_{n \geq 1}nT_nz_{\infty}^{-n-1}+T_0z_{\infty}^{-1}
-\sum_{n \geq 1}\frac{\partial F}{\partial T_n}z_{\infty}^{n-1}
\right\} dz_{\infty}~~~~~~\mbox{around}~~p_{\infty},
\nonumber \\
dS &=& \left\{
-\sum_{n \geq 1}n\bar{T}_n\tilde{z}_{\infty}^{-n-1}-T_0\tilde{z}_{\infty}^{-1}
-\sum_{n \geq 1}\frac{\partial F}{\partial \bar{T}_n}\tilde{z}_{\infty}^{n-1}
\right\} d\tilde{z}_{\infty}~~~~~~\mbox{around}~~\tilde{p}_{\infty}.
\label{expansion of dS-2}
\end{eqnarray}

          An interesting class of solutions of the Whitham hierarchy
are solutions that enjoy the  homogeneity condition
\begin{eqnarray}
\sum_{i=1}^{N-1}a_i\frac{\partial F}{\partial a_i}
+\sum_{n \geq 0}T_{n}\frac{\partial F}{\partial T_n}
+\sum_{n \geq 1}\bar{T}_{n}\frac{\partial F}{\partial \bar{T}_n}
=2F.
\label{homogeneity}
\end{eqnarray}
In this case, $dS$, which is introduced by eq.(\ref{dS-2}),
has a form analogous to (\ref{decomposition of dS-1})
\begin{eqnarray}
dS=
\sum_{i=1}^{N-1}a_idz_i
+\sum_{n \geq 0}T_{n}d\Omega_{\infty,n}
+\sum_{n \geq 1}\bar{T}_{n}d\tilde{\Omega}_{\infty,n}.
\label{dS-3}
\end{eqnarray}
Therefore one can reproduce $dS$  of (\ref{dS-1})
by simply setting
$T_1=-\bar{T}_1=1$ and other $T_A$-variables zero.
The pre-potential $\cal F$ of $N=2$ supersymmetric
$SU(N)$ Yang-Mills theory
is now given by
$\cal F$ $=2 \pi \sqrt{-1} F$.
Let us prove that  (\ref{dS-3}) is consistent with (\ref{expansion of dS-2}).
We first notice that eqs. (\ref{dS-2}) and (\ref{F}) make it possible
to express the coefficents in (\ref{expansion-1})
and (\ref{expansion-2}) in terms of $F$.
As for the holomorphic differentials $dz_i$
let us write their local expansion as
\begin{eqnarray}
 dz_i &=& -\sum_{m \geq 1} \sigma_{i,m}z_{\infty}^{m-1}dz_{\infty}~~~~
\mbox{around}~~p_{\infty},
\nonumber \\
dz_i &=& -\sum_{m \geq 1} \bar{\sigma}_{i,m}\tilde{z}_{\infty}^{m-1}
d\tilde{z}_{\infty}~~~~
\mbox{around}~~\tilde{p}_{\infty},
\label{expansion-3}
\end{eqnarray}
where $1 \leq i \leq N-1$. Then, by using eqs.(\ref{dS-2}) and (\ref{F}),
we can also express the coefficients in  (\ref{expansion-3}) in terms
of $F$. Inserting these local expansions into the R.H.S of (\ref{dS-3}),
and recalling the homogeneity of $F$ (\ref{homogeneity}),
one can reproduce (\ref{expansion of dS-2}).

      For this homogeneous solution,  we can write the $F$-function
in terms of the $\beta$-periods of the differentials
$dz_i, d\Omega_{\infty,n}$ and $d\Omega_{\infty, n}$ as follows.
First notice that by homogeneity relation (\ref{homogeneity}) and
definition of $F$-function (\ref{F}),  $F$ can be written
as a sum of periods and residues of $dS$
\begin{eqnarray}
F&=&
\frac{1}{2}\left(
\sum_{i=1}^{N-1}a_i\frac{\partial F}{\partial a_i}
+\sum_{n \geq 0}T_{n}\frac{\partial F}{\partial T_n}
+\sum_{n \geq 1}\bar{T}_{n}\frac{\partial F}{\partial \bar{T}_n} \right)
\nonumber \\
&=&
\frac{1}{2}\left\{
\sum_{i=1}^{N-1}\frac{a_i}{2 \pi \sqrt{-1}}\oint_{\beta_i}dS
-\sum_{n \geq 1}T_{n}res_{p_{\infty}}z_{\infty}^{-n}dS  \right.
\nonumber \\
&&~~\left.
-\sum_{n \geq 1}\tilde{T}_{n}
       res_{\tilde{p}_{\infty}}\tilde{z}_{\infty}^{-n}dS
-T_0 \left( res_{p_{\infty}}\ln z_{\infty}dS
     - res_{\tilde{p}_{\infty}}\ln \tilde{z}_{\infty}dS \right) \right\}.
\end{eqnarray}
Since $dS$ has the form given in (\ref{dS-3}),
this expression of $F$ can be evaluated further to become
the aforementioned form. Eventually, we obtain
\begin{eqnarray}
F&=&
\frac{1}{4 \pi \sqrt{-1}}\sum_{i,j=1}^{N-1}\tau_{i,j}a_ia_j
+\sum_{i=1}^{N-1}a_i
    \left\{ \sum_{k \geq 1}
             \left( \sigma_{i,k}T_k+\bar{\sigma}_{i,k}\bar{T}_k \right)
                 +\bar{\sigma}_{i,0}T_0 \right\}
\nonumber \\
&&+
\frac{1}{2}\sum_{k,l \geq 1}q_{k,l}T_kT_l
  +\sum_{k,l \geq 1}r_{k,l}T_k\bar{T}_l
     +\frac{1}{2}\sum_{k,l \geq 1}\bar{q}_{k,l}\bar{T}_k\bar{T}_l
       + \frac{1}{2}\bar{r}_{0,0}T_0^2
\nonumber \\
&&+
T_0 \sum_{k \geq 1}
            \left( r_{k,0}T_k+\bar{r}_{0,k}\bar{T}_k \right),
\label{homogeneous F}
\end{eqnarray}
where the quantities $\tau_{i,j},\bar{\sigma}_{i,0}$ and $\bar{r}_{0,0}$
are introduced as
\begin{eqnarray}
\tau_{i,j}&=&\oint_{\beta_i}dz_j~~~~~~(1 \leq i,j \leq N-1),
\nonumber \\
\bar{\sigma}_{i,0}&=&
\frac{1}{2 \pi \sqrt{-1}}\oint_{\beta_i}d\Omega_{\infty, 0}
{}~~~~(1 \leq i \leq N-1) ,
\nonumber \\
\bar{r}_{0,0}&=&
- res_{p_{\infty}}\ln z_{\infty}d\Omega_{\infty, 0}
+ res_{\tilde{p}_{\infty}}\ln \tilde{z}_{\infty}d\Omega_{\infty, 0},
\label{period}
\end{eqnarray}
and other quantities in (\ref{homogeneous F}) are those appearing
in expansions (\ref{expansion-1}),(\ref{expansion-2}) and (\ref{expansion-3}).

\section{}

   Now we will discuss the relation between this homogeneous solution of
the Whitham hierarchy (\ref{FFM-2}) and the Toda lattice hierarchy.
Let us denote Toda lattice time variables  by
$t=(t_1,t_2,\cdots)$ , $\bar{t}=(\bar{t}_1,\bar{t}_2,\cdots)$ and
$n$  ($n \in Z$). It is  well-known \cite{Toda curve} that
the Toda lattice hierarchy has a quasi-periodic solution associated
with a hyperelliptic curve $C$ as in (\ref{curveC}). In fact, this is
a dimensionally reduced solution --- in the lowest sector of the
hierarchy, the solution gives an $N$-periodic solution of the
Toda chain rather than the two-dimensional Toda field equations
\footnote{This dimensional reduction is irrelevant in our discussion.}.
Since the flows of the Toda lattice hierarchy do not change this curve
itself, the moduli parameters $u$ are invariants (integrals of motion)
of these solutions of the Toda lattice hierarchy.

       We  introduce an additional new time variables
$\theta=(\theta_1,\cdots,\theta_{N-1})$ into this solution.
This is achieved by modifying the associated Baker-Akhiezer function
\cite{Toda curve} as
\begin{eqnarray}
&&
\Psi(p;t,\bar{t},n,\theta)
\nonumber \\
&&=
\exp
\left\{
-n \int^p d\Omega_{\infty,0}
+\sum_{n \geq 1}t_n \int^p d\Omega_{\infty, n}
+\sum_{n \geq 1}\bar{t}_n \int^p d\tilde{\Omega}_{\infty, n}
+\sqrt{-1} \sum_{i=1}^{N-1} \theta_i \int^p dz_i \right\}
\nonumber \\
&&
\times
\frac{\vartheta
      \left( z(p)-z(D)+\Delta-n(z(p_{\infty})-z(\tilde{p}_{\infty}))
            +\sum_{n \geq 1}t_n \sigma_n
              +\sum_{n \geq 1}\bar{t}_n \bar{\sigma}_n
               +\frac{1}{2 \pi}\sum_{i=1}^{N-1}\theta_i \tau_i \right) }
     { \vartheta
       \left( z(p_{\infty})-z(D)+\Delta-n(z(p_{\infty})-z(\tilde{p}_{\infty}))
            +\sum_{n \geq 1}t_n \sigma_n
              +\sum_{n \geq 1}\bar{t}_n \bar{\sigma}_n
               +\frac{1}{2 \pi}\sum_{i=1}^{N-1}\theta_i \tau_i \right) }
\nonumber \\
&&
\times
\frac{\vartheta
         \left( z(p_{\infty})-z(D)+\Delta \right) }
     { \vartheta
         \left( z(p)-z(D)+\Delta \right) }.
\label{PSI}
\end{eqnarray}
$"z"$ is the Abel mapping, that is,
$z(p)=^{T}\!(z_1(p),\cdots,z_{N-1}(p))$ where $z_i(p)=\int^pdz_i$.
$D$ is a positive divisor of degree $N-1$ and $\Delta$ is the Riemann
constant. $\sigma_l$, $\bar{\sigma}_l$ ($l \geq 1$) and $\tau_{i}$
$(1 \leq i \leq N-1)$ are $N-1$ dimensional complex vectors whose
components are given by $\sigma_l=^T(\sigma_{l,1},\cdots,\sigma_{l,N-1})$,
$\bar{\sigma}_l=^T(\bar{\sigma}_{l,1},\cdots,\bar{\sigma}_{l,N-1})$,
and
$\tau_{i}=^T(\tau_{i,1},\cdots,\tau_{i,N-1})$.
Notice that Baker-Akhiezer function $\Psi$ (\ref{PSI})
reduces to that of the ordinary Toda lattice as $\theta  \rightarrow 0$.
For non-zero values of $\theta$, $\Psi$ is quasi-periodic
along the $\beta$-cycles,
\begin{eqnarray}
\Psi(p;t,\bar{t},n,\theta) \stackrel{\beta_i}{\longmapsto}
e^{\sqrt{-1}\theta_i}\Psi(p;t,\bar{t},n,\theta),
\end{eqnarray}
and  periodic along the $\alpha$-cycles.
Hence the Baker-Akhiezer function $\Psi$
can also be regarded as a section of
a flat line bundle over $C$  now labeled by
$\theta$-variables.
We also notice that this solution of the (generalized)
Toda lattice is $N$-periodic (with respect to $n$).
One can check the following periodicity at the level
of Baker-Akhiezer function
\begin{eqnarray}
\Psi(p;t,\bar{t},n+N,\theta)
= \mbox{const.}~~ h(p)~ \Psi(p;t,\bar{t},n,\theta) .
\end{eqnarray}

          One may introduce the $\tau$-function of this solution
following the prescription of Toda lattice hierarchy
\cite{Toda lattice hierarchy}. In particular the local expansions
of $\Psi$ around $p_{\infty}$ and $\tilde{p}_{\infty}$ will be
described by the $\tau$-function :
In a neighborhood of $p_{\infty}$
\begin{eqnarray}
\Psi(p;t,\bar{t},n,\theta) &=&
z_{\infty}^{-n}
e^{\sum_{l \geq 1}t_lz_{\infty}^{-l}}
\frac{\tau(t-\mbox{$[$} z_{\infty} \mbox{$]$}, \bar{t},n,\theta)}
      {\tau(t, \bar{t},n,\theta) },
\label{def of tau1}
\end{eqnarray}
and in a neighborhood of $\tilde{p}_{\infty}$
\begin{eqnarray}
\Psi(p;t,\bar{t},n,\theta) &=&
\tilde{z}_{\infty}^{n}
e^{\sum_{l \geq 1}\bar{t}_l\tilde{z}_{\infty}^{-l}}
\frac{\tau(t,\bar{t}-\mbox{$[$} \tilde{z}_{\infty} \mbox{$]$},n+1,\theta)}
      {\tau(t, \bar{t},n,\theta) },
\label{def of tau}
\end{eqnarray}
where
$\mbox{$[$} z_{\infty} \mbox{$]$}
=(\frac{z_{\infty}}{1},\frac{z_{\infty}^2}{2},\frac{z_{\infty}^3}{3},\cdots)$.

       By matching this expression of $\Psi$ with (\ref{PSI}),
we can write down the $\tau$-function in terms of the theta function
and the coefficients of local expansion of $dz_i$, etc.
We thus obtain the following expression of the $\tau$ function :
\begin{eqnarray}
&&\tau(t,\bar{t},n,\theta)  \nonumber \\
&&=e^{\hat{F}(t,\bar{t},n.\theta)} \nonumber \\
&&~~\times
 \vartheta
         \left( -(n-1)z(p_{\infty})+nz(\tilde{p}_{\infty})-z(D)+\Delta
            +\sum_{n \geq 1}t_n \sigma_n
              +\sum_{n \geq 1}\bar{t}_n \bar{\sigma}_n
               +\frac{1}{2 \pi}\sum_{i=1}^{N-1}\theta_i \tau_i \right),
\nonumber \\
{}~~~~~~~
\label{tau}
\end{eqnarray}
where $\hat{F}$ is a polynomial of second degree in
$t,\bar{t},n$ and $\theta$ given by
\footnote{In (\ref{hatF}) $d_k (k \geq 1)$ and $\bar{d}_k (k \geq 0)$
are the constants irrelevant to our
discussion and we do not describe their explicit forms.}
\begin{eqnarray}
\hat{F}(t,\bar{t},n,\theta)&=&
\frac{1}{2}\sum_{k,l \geq 1}q_{k,l}t_kt_l
  +\sum_{k,l \geq 1}r_{k,l}t_k\bar{t}_l
     +\frac{1}{2}\sum_{k,l \geq 1}\bar{q}_{k,l}\bar{t}_k\bar{t}_l
     -\frac{1}{4 \pi \sqrt{-1}}\sum_{i,j=1}^{N-1}\tau_{i,j}\theta_i \theta_j
\nonumber \\
&&+\frac{n(n-1)}{2}\bar{r}_{0,0}
  +\sqrt{-1}\sum_{i=1}^{N-1} \theta_i
    \left\{ \sum_{k \geq 1}
             \left( \sigma_{i,k}t_k+\bar{\sigma}_{i,k}\bar{t}_k \right)
                 -n\bar{\sigma}_{i,0} \right\}
\nonumber \\
&&
-n \sum_{k \geq 1} r_{k,0}t_k
-(n-1) \bar{r}_{0,k}\bar{t}_k
+\sum_{k \geq 1}d_kt_k
+\sum_{k \geq 1}\bar{d}_k\bar{t}_k+\bar{d}_0n.
\nonumber \\
&&~~
\label{hatF}
\end{eqnarray}

           Note that the quadratic part of $\hat{F}$ (\ref{hatF})
has almost the same form as  $F$ in (\ref{homogeneous F}).
This is not accidental. In fact it turns out that the Whitham
hierarchy in the previous section gives modulation equations
of the aforementioned quasi-periodic solution of the (generalized)
Toda lattice hierarchy. Let us explain this relation in more detail
following Bloch and Kodama \cite{modulation}
\footnote{Their Whitham hierarchy for the Toda lattice is
different from ours, but the idea of deriving modulation
equations is the same.}.
For this purpose we introduce "slow" time variables
\footnote{$\epsilon$ is arbitary small. So these variables
are small relative to the (generalized) Toda time variables.}
$T_l,\bar{T}_l (l \geq 1),T_0$ and $a_i (1 \leq i \leq N-1)$ by
\begin{eqnarray}
T_l=\epsilon t_l~,~~\bar{T}_l=\epsilon \bar{t}_l~,~~
T_0=-\epsilon n~,~~a_i=\sqrt{-1} \epsilon \theta_i.
\label{slow variables}
\end{eqnarray}
and consider the asymptotics of Baker-Akhiezer function
(\ref{PSI}) and $\tau$-function (\ref{tau}) as
$\epsilon \to 0$. In the slow time variables,
they turn out to have the following expression
\begin{eqnarray}
d \ln
\Psi \left(p; T/\!\epsilon ,
               \bar{T}/\!\epsilon, -T_0/\!\epsilon ,
                  a/\!\sqrt{-1}\epsilon  \right)
&=&
\epsilon^{-1} \sum_{n \geq 0} \epsilon^n
       dS^{(n)}(p;T,\bar{T},T_0,a) ,
\nonumber \\
\ln \tau \left( T/\!\epsilon ,
               \bar{T}/\!\epsilon, -T_0/\!\epsilon ,
                  a/\!\sqrt{-1}\epsilon  \right)
&=&
\epsilon^{-2} \sum_{n \geq 0} \epsilon^{n}
       F^{(n)}(T,\bar{T},T_0,a).
\label{dispersionless}
\end{eqnarray}
The leading-order terms $dS^{(0)}$ and $F^{(0)}$
are given by the same expressions  as $dS$ in  (\ref{dS-3})
and $F$ in (\ref{homogeneous F}) respectively.
We now allow the moduli parameters $u$ to depend
on the slow variables,
\begin{eqnarray}
u_{k}=u_{k}(T,\bar{T},T_0,a).
\end{eqnarray}
In other words, the hyperelliptic curve $C$ can now
$''$ slowly $''$ vary as $C = C(T,\bar{T},T_0,a)$.
We further require that this $''$modulated $''$ quasi-periodic
wave be still a solution of the (generalized) Toda lattice
hierarchy. This induces a system of differential equations
(modulation equations) to the moduli parameters. By the
theory of Whitham hierarchies \cite{Whitham hierarchy2},
\cite{D-L}, \cite{Whitham hierarchy}, these modulation
equations turn out to be nothing but our Whitham hierarchy
(\ref{FFM-2}).

\section{}

    So far, we have considered  the integrable structure
of $N=2$ supersymmetric Yang-Mills theory
by embedding the system into the Whitham hierarchy (\ref{FFM-2}).
The hierarchy is constructed by adding the additional $T$- and
$\bar{T}$-flows to the flows generated by the holomorphic differentials.
Meanwhile, the time variables $a$ of the latter flows, along with their
dual variables $a_D$, constitute the units of spectrum in the theory.
Though all these flows can be identified with those of the (generalized)
Toda lattice, physical roles of the additional ones are not clear.
In this final section we discuss their physical implication.
In particular we  show that the solution of the Whitham hierarchy
that we have considered  satisfies the same Virasoro constraints as appear in
topological string theory.

For this purpose we define the following function Q
by using the differential $dS$  introduced in (\ref{dS-2}) ,
\begin{eqnarray}
Q=\frac{dS}{dh}.
\label{Q}
\end{eqnarray}
Since $S(p)=\int^pdS$ is  multi-valued,  $Q$ can also be
multi-valued in general. Actually, it is known
\cite{Whitham hierarchy2} that a class of solutions of the
Whitham hierarchy are characterized by geometric conditions
concerning single-valuedness and regularity of $Q$.
For an example one can determine a solution of the Whitham
hierarchy by the condition that $Q$ be single-valued and
have no singular points other than $p_{\infty}$ and
$\tilde{p}_{\infty}$.  This is indeed the solution that
we have considered in the preceding sections. Let us now
impose a stronger condition that this single-valued $Q$
be regular at $\tilde{p}_{\infty}$, too.
This regularity condition forces $\forall \bar{T}_k=0$.
Furthermore, by the single-valuedness, it follows that
\begin{eqnarray}
res_{p_{\infty}}h^{n+1}QdS=0
\label{Virasoro1}
\end{eqnarray}
for $\forall n \geq -1$.
This is a consequence of the Riemann bilinear relation.
Since we already know how the differentials
$h^{n+1}QdS$ behave around $p_{\infty}$,
eqs.(\ref{Virasoro1}) can be rephrased as constraints
on the $F$-function. They are given by
\begin{eqnarray}
\sum_{k \geq 1}kT_k\frac{\partial F}{\partial T_{k+nN}}
+\frac{1}{2}\sum_{k+l=nN}\frac{\partial F}{\partial T_k}
            \frac{\partial F}{\partial T_l}
+\frac{1}{2}\sum_{k+l=N}klT_kT_l\delta_{n+1,0}=0
\label{Virasoro2}
\end{eqnarray}
for $\forall n \geq -1$.
Notice that
we can make $F$ independent of $T_{kN} (\forall k\geq 1)$ in
(\ref{Virasoro2}) without any loss of generality. This is
because the corresponding generators become exact forms,
\begin{eqnarray}
d\Omega_{\infty,kN}=dh^k~~ .
\end{eqnarray}
Hence constraints (\ref{Virasoro2}) now become  precisely same as
those on the free energy of the $A_{N-1}$ topological string
(in spherical limit) \cite{dKP},\cite{singularity}.
In this topological string theory, time variables such as $T_n$ play
the role of coupling constants of the chiral primary fields or their
gravitational descendants.

     From this observation  it would be a very fascinating idea to
interpret the meromorphic differentials $d\Omega_{\infty,n}$
$(n \geq 1)$ as the chiral primary fields or their gravitational
descendants of  the $A_{N-1}$ topological string. It should  be
noticed that one can also give the same interpretation on the
differentials $d\tilde{\Omega}_{\infty,k}$$(k \geq 1)$.
So it might be possible to interpret  N=2 supersymmetric
Yang-Mills theory as a coupled system of two topological string
models. As we have mentioned above, the corresponding solution
of the Whitham hierarchy is characterized by the condition
that $Q$ be single-valued and have singularities at most at
$p_{\infty}$ and $\tilde{p}_{\infty}$. Again by the Riemann
bilinear relations, this condition can be restated as
\begin{eqnarray}
res_{p_{\infty}}h^{n+1}QdS
+ res_{\tilde{p}_{\infty}}h^{n+1}QdS
=0
\label{Virasoro3}
\end{eqnarray}
for $\forall n$.
We can rewrite  (\ref{Virasoro3}) as constraints on the
$F$-function. They turn out to be
\begin{eqnarray}
\sum_{k \geq 1}kT_k\frac{\partial F}{\partial T_{k+nN}}
+\frac{1}{2}\sum_{k+l=nN}\frac{\partial F}{\partial T_k}
    \frac{\partial F}{\partial T_l}
&=&
\Lambda^{2nN}\left\{
\sum_{k \geq 1}(k+nN)\bar{T}_{k+nN}\frac{\partial F}{\partial \bar{T}_{k}}
+\frac{1}{2}\sum_{k+l=nN}kl\bar{T}_{k}\bar{T}_{l} \right\},
\nonumber \\
\sum_{k \geq 1}k\bar{T}_k\frac{\partial F}{\partial \bar{T}_{k+nN}}
+\frac{1}{2}\sum_{k+l=nN}\frac{\partial F}{\partial \bar{T}_k}
\frac{\partial F}{\partial \bar{T}_l}
&=&
\Lambda^{2nN}\left\{
\sum_{k \geq 1}(k+nN)T_{k+nN}\frac{\partial F}{\partial T_{k}}
+\frac{1}{2}\sum_{k+l=nN}klT_{k}T_{l} \right\},
\nonumber \\
&&~~~~
\end{eqnarray}
where $n \geq 0$.

\subsection{Acknowledgements}

The second author (K.T.) would like to thank A. Gorsky,
S. Kharchev, A. Marshakov, A. Mironov, and A. Morozov
for fruitful discussions. The idea to connect the
hyperelliptic cirve $C$ with the Toda lattice was
inspired by this Moscow group.


\end{document}